\documentclass[pra,showpacs,amsmath,amssymb]{revtex4}
\usepackage{graphicx}
\usepackage{bm}
\usepackage{amsmath,amssymb}
\begin{document}

\title{Superposition of macroscopically distinct states
means large multipartite entanglement}

\author{Tomoyuki Morimae}
\email{morimae@asone.c.u-tokyo.ac.jp}
\affiliation{
Department of Basic Science, University of Tokyo, 
3-8-1 Komaba, Tokyo 153-8902, Japan}
\affiliation{
Laboratoire Paul Painlev\'e, Universit\'e Lille 1,
59655 Villeneuve d'Ascq C\'edex, France
}
\date{\today}
            
\begin{abstract}
We show relations between superposition of macroscopically
distinct states and entanglement.
These relations lead to 
the important conclusion that
if a state contains superposition of macroscopically
distinct states, the state also contains
large multipartite entanglement in terms of several
measures.  
Such multipartite entanglement property also
suggests that
if a state contains superposition of macroscopically
distinct states, a measurement on a single particle 
drastically changes the state of macroscopically
many other particles, as in the case of the $N$-qubit GHZ state.
\end{abstract}
\maketitle  
\section{Introduction}
Superposition of macroscopically distinct states 
is one of the most 
fundamental characteristics in 
quantum physics,
which has been attracting much attentions since
the birth of quantum physics~\cite{Schrodinger}. 
On the experimental side,
such superposition has been explored in
various many-body systems, such as
magnetic materials~\cite{magneticMQT} and
trapped ions~\cite{iontrapcat}. 
On the theoretical side, 
useful quantities and methods have been proposed
in order to quantitatively investigate 
such superposition~\cite{Leggett,mermin,ciraccat,ciraccat2,delft,SM,SM05}.
For example,
Leggett introduced 
the criterion, ``disconnectivity",
and studied superposition of macroscopically distinct states
in superconducting systems~\cite{Leggett}.
Mermin proposed a many-partite Bell-like inequality, and showed
that it is exponentially violated for
the $N$-qubit GHZ state
$|\mbox{GHZ}\rangle
\equiv(|0^{\otimes N}\rangle+|1^{\otimes N}\rangle)
/\sqrt{2}$, which is a typical example of superposition
of macroscopically distinct states~\cite{mermin}.

Entanglement is the other important property
of quantum states. After being 
introduced by Schr\"odinger~\cite{Schrodinger},
a great deal of research has been performed on the 
physical and mathematical characteristics of
entanglement itself~\cite{entanglement} and 
on the applications of 
entanglement to 
condensed matter physics~\cite{Amico}.
Nowadays, entanglement is also known as
a crucial resource for quantum information processing~\cite{Nielsen}.

Since both superposition of macroscopically distinct states
and entanglement represent ``quantumness" of physical systems,
it is reasonable to expect that
there are some
fundamental relations between them.
However, to the author's knowledge, such relations have so far eluded us.
The purpose of this Rapid Communication is to show several
relations between superposition of macroscopically
distinct states and entanglement.
From these relations, we obtain the important conclusion
that
a state which contains superposition of macroscopically
distinct states also contains large multipartite entanglement
in terms of several measures, such as the localizable
entanglement~\cite{localizable}, 
the distance-like measure of entanglement~\cite{Vedral1}, 
and multipartite entanglement defined through various 
bipartitions.

Furthermore, such large multipartite entanglement 
property also
leads to another interesting consequence.
As is well known,
the projective measurement 
$\{|0\rangle\langle0|,|1\rangle\langle1|\}$ 
on a single particle 
of the $N$-qubit GHZ state
drastically changes  
the state of other $N-1$ particles. 
Since the $N$-qubit GHZ state is a typical example of 
superposition of macroscopically distinct states,
it is legitimate to think that
other states which contain
superposition of macroscopically distinct states
also have such property.
In fact, we will show that 
it is the case.

\section{Index $p$}
As the criterion of superposition of macroscopically distinct states,
we use index $p$~\cite{SM}.
Let us consider an $N$-site lattice
($1\ll N<\infty$) where
the dimension of the Hilbert space on each site is 
an $N$-independent constant,
such as a chain of $N$ spin-1/2 particles.
Since we are interested in the macroscopic
properties of the system,
we use two symbols, $O$ and $o$, in order to describe 
asymptotic behaviors of a function $f(N)$
in the thermodynamic limit $N \to \infty$:
$f(N)=O(N^k)$ means
$\lim_{N\to\infty}[f(N)/N^k]=\mbox{const.}\neq0$
and
$f(N)=o(N^k)$ means
$\lim_{N\to\infty}[f(N)/N^k]=0$.

For a given pure state $|\psi\rangle$, 
the index $p$ $(1\leq p \leq 2)$ is defined by
\begin{eqnarray*}
\max_{\hat{A}}C(\hat{A},\hat{A},|\psi\rangle)=O(N^p),
\end{eqnarray*}
where
$C(\hat{X},\hat{Y},|\psi\rangle)
\equiv
\langle\psi|\hat{X}\hat{Y}|\psi\rangle
-\langle\psi|\hat{X}|\psi\rangle
\langle\psi|\hat{Y}|\psi\rangle$
is the correlation,
and the maximum 
is taken over all Hermitian additive operators 
$\hat{A}$.
Here, an additive operator 
$\hat{A}
=\sum_{l=1}^N\hat{a}(l)$
is a sum of local operators $\{\hat{a}(l)\}_{l=1}^N$,
where $\hat{a}(l)$
is a local operator acting on site $l$.
For example, if the system is a chain 
of $N$ spin-$1/2$ particles,
$\hat{a}(l)$ is a linear combination
of three Pauli operators, 
$\hat{\sigma}_x(l),\hat{\sigma}_y(l),\hat{\sigma}_z(l)$, 
and the identity operator $\hat{1}(l)$ acting on site $l$.
In this case, the $x$-component of the total
magnetization 
$\hat{M}_x\equiv\sum_{l=1}^N\hat{\sigma}_x(l)$
and the $z$-component of the total staggard magnetization
$\hat{M}_z^{st}\equiv\sum_{l=1}^N(-1)^l\hat{\sigma}_z(l)$
are, for example, additive operators.
The index $p$ takes the minimum value 1 for any 
product state
$\bigotimes_{l=1}^N|\phi_l\rangle$,
where $|\phi_l\rangle$ is a state of site $l$
(this means that $p>1$ is an entanglement witness
for pure states). 
If $p$ takes the maximum value 2,
the state contains superposition of macroscopically 
distinct states,
because in this case a Hermitian additive operator  
has a ``macroscopically large" fluctuation
in the sense that
the relative fluctuation does not vanish in
the thermodynamic limit:
$\lim_{N\to\infty}
\big[C(\hat{A},\hat{A},|\psi\rangle)^{1/2}/N\big]
\neq0$,
and because the fluctuation of an observable
in a pure state means the existence of a superposition
of eigenvectors of that observable
corresponding to different eigenvalues.
For example, the $N$-qubit GHZ state 
$|\mbox{GHZ}\rangle
\equiv(|0^{\otimes N}\rangle+|1^{\otimes N}\rangle)/\sqrt{2}$, 
which obviously contains
superposition of macroscopically distinct states,
has $p=2$, since
$C(\hat{M}_z,\hat{M}_z,|\mbox{GHZ}\rangle)=N^2$.

\section{Bipartite entanglement}
Let us first briefly
examine relations of $p$ to bipartite entanglement.
One of the simplest ways of quantifying bipartite
entanglement in a quantum many-body system
is to divide the
total system into two equal-size subsystems,
e.g., dividing $\{1,2,...,N\}$ into
$\{1,2,...,N/2\}$ and $\{N/2+1,...,N\}$,
and calculate the von Neumann entropy 
$E$ ($0\le E\le N/2$) 
of the
reduced density operator of one subsystem.
For translationally invariant systems, which
are ubiquitous in condensed matter physics,
this is a good way of evaluating
bipartite entanglement.
It is easy to show that the BEC of magnons
(or $N/2$-Dicke state):
$\Big(\sum_{l=1}^N|1\rangle_l\langle0|\Big)^{N/2}
|0^{\otimes N}\rangle$
has $p=2$ whereas the entanglement entropy $E$ 
is as small as $O(\log N)$ $(\ll N/2)$.
Furthermore, it is immediate to show
$E=\log2$ for the $N$-qubit GHZ state which has $p=2$.
Therefore, superposition of macroscopically
distinct states
does not necessarily mean large bipartite entanglement
in terms of this measure.

The other way of quantifying bipartite
entanglement in a quantum many-body system
is to choose two representative sites
and evaluate 
the concurrence~\cite{concurrence} between them.
This approach is often taken
in the study of entanglement behavior
at quantum critical points.
It is easy to calculate that the concurrence between any two sites
of the $N$-qubit GHZ state is 0, 
whereas that of the BEC of magnons  
is $O(1/N)$
(note that $2/N$ is the maximum value of the concurrence 
for homogeneous systems~\cite{Koashi}).
Since these two states both have $p=2$, 
there is no direct connection between
$p$ and concurrence (and hence the entanglement 
of formation).

In short, we conclude that there is no significant
relation between superposition of macroscopically
distinct states and bipartite entanglement.
In particular, superposition of macroscopically distinct states
does not necessarily mean
large bipartite entanglement.
In the followings, we next consider multipartite entanglement.

\section{Localizable entanglement}
The localizable entanglement between
two sites is defined by the maximum
amount of entanglement
that can be localized in these two sites, on average,
by doing local measurements on other sites~\cite{localizable}.

We can show that if $p=2$, the state has large
multipartite entanglement in the sense that
macroscopically many $(O(N^2))$ pairs of sites
have non-vanishing amount ($O(1)$) of localizable entanglement.
In order to show it, let 
$S\equiv\{1,2,...,N\}$ be the set of all sites,
$|\psi\rangle$ be the state of the total system, 
and $\hat{A}=\sum_{l=1}^N\hat{a}(l)$ be
an additive operator.
Let us define two subsets
\begin{eqnarray*}
R_1&\equiv&
\{(l,l')\in S\times S~|~C(\hat{a}(l),\hat{a}(l'),|\psi\rangle)=O(1)\}\\
R_2&\equiv&
\{(l,l')\in S\times S~|~C(\hat{a}(l),\hat{a}(l'),|\psi\rangle)=o(1)\}
\end{eqnarray*}
of $S\times S$.
In other words,
$R_1$ is the set of pairs such that the correlation
persists in the thermodynamic limit,
and $R_2$ is that of others. 
Let us
assume that the number $|R_1|$ of elements of $R_1$
is $o(N^2)$ for any $\hat{A}$.
Then,
\begin{eqnarray*}
C(\hat{A},\hat{A},|\psi\rangle)&=&
\sum_{l=1}^N\sum_{l'=1}^NC(\hat{a}(l),\hat{a}(l'),|\psi\rangle)\\
&=&\Big[\sum_{(l,l')\in R_1}
+\sum_{(l,l')\in R_2}\Big]
C(\hat{a}(l),\hat{a}(l'),|\psi\rangle)\\
&\le& o(N^2)O(1)+O(N^2)o(1)=o(N^2)
\end{eqnarray*}
for any $\hat{A}$,
which means $p<2$.
Therefore, if $p=2$, $|R_1|=O(N^2)$ for an $\hat{A}$.

It is known that
the localizable entanglement between two sites
is lower bounded
by the maximum correlation between these two sites~\cite{localizable}.
Therefore, the above result
means that 
if $p=2$, 
macroscopically
many $(O(N^2))$ pairs have
finite $(O(1))$ amount of
localizable entanglement in the thermodynamic
limit.
In this sense, a state which contains 
superposition of macroscopically distinct states
has large multipartite entanglement.

\section{Distance-like measure of entanglement}
In Ref.~\cite{Vedral1}, a measure of multipartite entanglement
for $\hat{\rho}$ was defined by
\begin{eqnarray*}
E_D(\hat{\rho})\equiv
\min_{\hat{\sigma}}D(\hat{\rho},\hat{\sigma}).
\end{eqnarray*}
Here, $D$ is a distance in the Hilbert space
and the minimum is taken over all separable state
$\hat{\sigma}\equiv
\sum_i\lambda_i\bigotimes_{l=1}^N|\phi_l^{(i)}\rangle
\langle\phi_l^{(i)}|$,
where $0\le\lambda_i\le1$, $\sum_i\lambda_i=1$,
and $|\phi_l^{(i)}\rangle$ is a state of site $l$.

In order to clarify the relation between superposition
of macroscopically distinct states and $E_D$, it is necessary
to consider mixed states.
However,
$p=2$ is not a witness
of superposition of macroscopically distinct states
in mixed states,
since a fluctuation is not necessarily equivalent
to a coherence in mixed states (for example, consider
the state
$|0^{\otimes N}\rangle\langle0^{\otimes N}| 
+|1^{\otimes N}\rangle\langle1^{\otimes N}|$,
which has macroscopically large fluctuation
but no coherence). 

Index $q$, which was introduced in Ref.~\cite{SM05}, is a criterion 
of superposition of macroscopically distinct states
in mixed states.
For a given many-body state $\hat{\rho}$,
index $q$ ($1\le q\le 2$) is defined by
\begin{eqnarray*}
\max\Big(N,\max_{\hat{A}}
\Big\|[\hat{A},[\hat{A},\hat{\rho}]]\Big\|_1\Big)
=O(N^q),
\end{eqnarray*}
where $\|\hat{X}\|_1\equiv
\mbox{Tr}\sqrt{\hat{X}^\dagger\hat{X}}$ is the 1-norm, and 
$\max_{\hat{A}}$ means the maximum over all
Hermitian additive operators $\hat{A}$.
As detailed in Ref.~\cite{SM05}, 
$q$ takes the minimum value 1 for any separable state,
and
if $q$ takes the maximum value 2,
the state contains superposition
of macroscopically distinct states.
In particular,
for pure states, $p=2\iff q=2$.

Let $\hat{\sigma}$ be a separable state
and $\hat{\rho}$ be a state having $q=2$.
Then,
\begin{eqnarray*}
O(1)&=&
\frac{
\big\|[[\hat{\rho},\hat{A}],\hat{A}]\big\|_1
-\big\|[[\hat{\sigma},\hat{A}],\hat{A}]\big\|_1
}{N^2}\\
&\le&
\frac{
\big\|[[\hat{\rho}-\hat{\sigma},\hat{A}],\hat{A}]\big\|_1}{N^2}
\le
4\|\hat{\rho}-\hat{\sigma}\|_1
\end{eqnarray*}
for an additive operator $\hat{A}$.
Therefore, if we choose the 1-norm as the distance $D$,
we obtain $E_D(\hat{\rho})=O(1)$, which means that if $q=2$
the state has persistent multipartite entanglement in the thermodynamic
limit.

The Bures distance and the relative entropy are often
used as the distance in $E_D$.
The Bures distance is defined by
$\sqrt{2(1-\|\sqrt{\hat{\rho}}\sqrt{\hat{\sigma}}\|_1)}$.
Then, we can show that
if $q=2$, $E_D(\hat{\rho})$ is as large as $O(1)$,
since
$1-\|\sqrt{\hat{\rho}}\sqrt{\hat{\sigma}}\|_1
=
(1-\|\sqrt{\hat{\rho}}\sqrt{\hat{\sigma}}\|_1^2)/
(1+\|\sqrt{\hat{\rho}}\sqrt{\hat{\sigma}}\|_1)
\ge\frac{1}{8}\|\hat{\rho}-\hat{\sigma}\|_1^2
=O(1)$~\cite{Nielsen}. 
On the other hand,
the relative entropy is defined by
$S(\hat{\rho}\|\hat{\sigma})\equiv\mbox{Tr}(
\hat{\rho}\log\hat{\rho}-\hat{\rho}\log\hat{\sigma})$.
By using the well-known inequality
$S(\hat{\rho}\|\hat{\sigma})\ge
\frac{1}{2}\|\hat{\rho}-\hat{\sigma}\|_1^2$~\cite{Ohya},
we again obtain
$E_D(\hat{\rho})\ge O(1)$ if $q=2$.

In summary, a state which contains superposition of macroscopically
distinct states also has
large multipartite entanglement in terms of
the distance-like measures of entanglement~\cite{robustness}.

\section{Multipartite entanglement defined through 
various bipartitions}
Another way of evaluating multipartite entanglement
in quantum many-body systems
is to consider various bipartitions
and evaluate bipartite entanglement among them.
For example,
Meyer and Wallach introduced the multipartite entanglement
measure
$2(1-\frac{1}{N}\sum_{l=1}^N\mbox{Tr}(\hat{\rho}_l^2))$
by considering all bipartition between a single site
and others,
where $\hat{\rho}_l$ is the reduced density operator
of site $l$~\cite{Meyer}.
Here, we consider similar multipartite entanglement.

Let $S\equiv\{1,2,...,N\}$ be the set of all sites
and $|\psi\rangle_S$ be the state of the total system.
Consider 
the Schmidt decomposition 
between site $l$ and other sites, i.e., $S-l$:
\begin{eqnarray*}
|\psi\rangle_S=\sqrt{\lambda_0}|\xi_0\rangle_l|\phi_0\rangle_{S-l}
+\sqrt{\lambda_1}|\xi_1\rangle_l|\phi_1\rangle_{S-l}.
\end{eqnarray*}
Then, let us factor 
out, if any, the common state $|\omega\rangle$ 
in $|\phi_0\rangle_{S-l}$ and $|\phi_1\rangle_{S-l}$ as
\begin{eqnarray*}
|\psi\rangle_S&=&\Big(
\sqrt{\lambda_0}|\xi_0\rangle_l|\eta_0\rangle_{S_1(l)}
+\sqrt{\lambda_1}|\xi_1\rangle_l|\eta_1\rangle_{S_1(l)}
\Big)|\omega\rangle_{S_2(l)}
\end{eqnarray*}
so that the number $|S_2(l)|$ of sites in the subsystem
$S_2(l)$ is maximum.
Here, 
\begin{eqnarray*}
|\phi_0\rangle_{S-l}&=&
|\eta_0\rangle_{S_1(l)}\otimes|\omega\rangle_{S_2(l)}\\
|\phi_1\rangle_{S-l}&=&
|\eta_1\rangle_{S_1(l)}\otimes|\omega\rangle_{S_2(l)}
\end{eqnarray*}
and $S_1(l)+S_2(l)=S-l$. 
Since the states of $l+S_1(l)$ is pure,
entanglement 
between $l$ and $S_1(l)$ is quantified 
by the entanglement entropy 
$E(l)$ $(0\le E(l)\le 1)$ 
as
$E(l)=-\lambda_0\log_2\lambda_0-\lambda_1\log_2\lambda_1$.

Let us consider the quantity 
$E_B$ 
($0\le E_B\le N$) defined by
\begin{eqnarray*}
E_B\equiv |\{l\in S~|~E(l)=O(1)~\mbox{and}~|S_1(l)|=O(N)\}|,
\end{eqnarray*}
where $|X|$ is the number of elements in the set $X$.
$E(l)=O(1)$ means that entanglement
between $l$ and $S_1(l)$
does not vanish in the thermodynamic limit.
$|S_1(l)|=O(N)$ means that site $l$ is entangling
with macroscopically many other sites.
Therefore,
$E_B$ is
the number of sites each of which is entangling
with macroscopically many other sites with non-vanishing
amount of entanglement.
For example, the cluster state~\cite{cluster}
has maximum entanglement in terms of $E_B=N$, since
it is easily confirmed that
$E(l)=1$ and $|S_1(l)|=N-1$ for any $l$.

We can show that
if $|\psi\rangle_S$ has $p=2$,
$E_B=O(N)$
(a proof is given in Appendix).
This means that a state which contains superposition
of macroscopically distinct states also has large multipartite
entanglement in the sense of $E_B=O(N)$.
Note that this also means the Meyer-Wallach's measure is large
if $p=2$.

\section{Effect of a measurement on a single particle}
The advantage of considering $E_B$
is that the effect of a measurement on a single particle becomes very clear.
Let us randomly choose a single site, say site $l$, from $S$.
From the above result, 
site $l$ satisfies
$E(l)=O(1)$ and $|S_1(l)|=O(N)$
with the non-vanishing probability $E_B/N=O(1)$ if $p=2$.
The projective measurement 
$\{|\xi_0\rangle\langle\xi_0|,|\xi_1\rangle\langle\xi_1|\}$
on site $l$
changes the state
of $S_1(l)$ into 
$|\eta_0\rangle$ 
or
$|\eta_1\rangle$, 
and this change is a ``drastic" one
since
(i) the information gain
through this measurement
is as large as $E(l)=O(1)$
and 
(ii) the state of 
macroscopically many $|S_1(l)|=O(N)$ sites are changed
by this measurement.
Therefore, we obtain the second main conclusion
that if a state contains superposition of macroscopically distinct states,
a measurement on a single site drastically
changes the state of macroscopically many other sites.

\section{Conclusion and Discussion}
In this Rapid Communication,
we have shown relations between 
superposition of macroscopically
distinct states and entanglement,
and concluded that
if a state contains superposition of macroscopically 
distinct states,
the state also contains large multipartite entanglement 
in terms of several measures.
We have also seen that if a state contains superposition
of macroscopically distinct states,
a projective measurement on a single particle
drastically changes the state of macroscopically many other particles.

Since there are infinitely many measures for multipartite entanglement, 
and each measure sees different features of quantum many-body states, 
it is unrealistic to expect 
that superposition of macroscopically distinct states means
large multipartite entanglement in terms of {\it any} measures.
Indeed, if we say a state has a large multipartite entanglement if macroscopically
many particles are genuinely entangled~\cite{Toth}, 
a superposition of macroscopically distinct states does not necessarily mean large 
multipartite entanglement, since a weak entanglement among macroscopically many particles is
not enough for a state to have a macroscopic superposition 
(for example, the W-state).
In the similar reasoning, we can conclude that a superposition of macroscopically
distinct states does not necessarily mean large multipartite entanglement
if the multipartite entanglement is defined by 
the minimum of the bipartite entanglement over all bipartitions~\cite{Pope}.

Furthermore, if we consider cluster states~\cite{cluster}, 
the discrepancy between
large multipartite entanglement and macroscopic superposition
becomes clear.
It is known that a multipartite Bell's inequality~\cite{Toth},
the Schmidt measure~\cite{hein}, 
and the geometric measure of entanglement~\cite{Barnum} take 
large values for cluster states, 
whereas $p=1$ for cluster states since they have no long-range
two-point correlations.
One of the reasons for this discrepancy seems to be the well known
fact that the ``quantumness" of cluster states is hidden in 
many-point correlations.
In order to gain insight into this, 
let us consider a simple example, the RVB state:
$|1,2\rangle|3,4\rangle...|N-1,N\rangle
+|2,3\rangle|4,5\rangle...|N,1\rangle$,
where $|i,j\rangle$ is the singlet between sites $i$ and $j$.
The RVB state is obviously the superposition of 
two macroscopically distinct VB states.
However, the maximum entanglement between nearest-neighbor
sites prohibits the existence of long-range two-point correlations in this state (i.e., entanglement
monogamy). Therefore, the RVB state has $p=1$. In spite of it,
the RVB state has large multipartite entanglement in terms of several measures
(for example, the measure $E_B$ considered in this paper is as large as $O(N)$).
It is known that in order to see quantum correlations in this state,
at least four-point correlations are required.

The detailed analysis of the discrepancy between large multipartite entanglement and
macroscopic superposition is, however, beyond the scope of
the present paper. It is an important subject of the future study.
 
Finally, let us briefly discuss the relation of our results to the
entanglement witness~\cite{Toth}.  Experimental detections of multipartite
entanglement is one of the most important subjects in today's quantum
many-body physics, and many detection methods, i.e., witnesses, have
been proposed~\cite{Toth}. Our results between superposition of macroscopically distinct states
and multipartite entanglement imply that an experimental detection of 
macroscopic superposition is also a witness of multipartite entanglement
in terms of several measures. An advantage of the detection of multipartite
entanglement through the detection of macroscopic superposition is that
it detects not only inseparability but also large multipartite entanglement.
Among various entanglement witnesses, in particular, 
indices $p$ and $q$ are closely related to the witness 
through ``collective measurements'',
such as the spin-squeezing parameter and the magnetic susceptibility~\cite{Toth}, 
since $p$ and $q$ are defined by using additive operators. 
Because of the uncertainty relation, the squeezing of one component of 
the total magnetization
leads to the large fluctuation of the other component. This large fluctuation
represents the macroscopic superposition and large multipartite entanglement.
On the other hand, the magnetic susceptibility is proportional to the fluctuation
of the magnetization. The multipartite entanglement properties of a 
many-body ground state often gives long-range two-point correlations 
and therefore a large fluctuation
of a component of the magnetization. 
This persists at sufficiently low temperature, and is
detected through the measurement of the magnetic susceptibility.

\acknowledgements
The author thanks A. Shimizu and Y. Matsuzaki for 
valuable discussions.
This work was partially supported by Japan Society for the
Promotion of Science.

\appendix*
\section{}
Let $|\psi\rangle$
be the state of the total system $S$,
and let us decompose $|\psi\rangle$ into 
a tensor product of inseparable states:
$|\psi\rangle=\bigotimes_{i=1}^r|\psi_i\rangle$,
where $|\psi_i\rangle$'s are inseparable states.
We denote the subsystem corresponding to
$|\psi_i\rangle$ by $Z_i$,
i.e., $Z_1+Z_2+...+Z_r=S$.
Let us also decompose an additive operator 
$\hat{A}=\sum_{l=1}^N\hat{a}(l)$
according to this partition
as $\hat{A}=\sum_{i=1}^r\hat{A}_i$, where
$\hat{A}_i\equiv\sum_{l\in Z_i}\hat{a}(l)$
is an operator acting on $Z_i$.
Then,
$C(\hat{A},\hat{A},|\psi\rangle)
=\sum_{i=1}^rC(\hat{A}_i,\hat{A}_i,|\psi_i\rangle)$,
which means that if $|\psi\rangle$ has $p=2$,
there exists at least one $|\psi_i\rangle$ which has $p=2$
and $|Z_i|=O(N)$.
Without loss of generality, we assume that
$|\psi_1\rangle$ has $p=2$ and
$|Z_1|=O(N)$.
Let us consider the Schmidt decomposition 
of $|\psi_1\rangle$
between a single site $l\in Z_1$ and the rest
of it $Z_1-l$:
\begin{eqnarray*}
|\psi_1\rangle=\sqrt{\lambda_0}|\xi_0\rangle_l|\eta_0\rangle_{Z_1-l}
+\sqrt{\lambda_1}|\xi_1\rangle_l|\eta_1\rangle_{Z_1-l}.
\end{eqnarray*}
Without loss of generality, we assume $\lambda_0\ge\lambda_1$.
Since $|\psi_1\rangle$ is inseparable by assumption,
$|\eta_0\rangle_{Z_1-l}$
and
$|\eta_1\rangle_{Z_1-l}$
have no common factor.
Let us define local operators on site $m$ $(m\in Z_1)$
as
$\hat{t}_x(m)\equiv
|\xi_0\rangle\langle\xi_1|+|\xi_1\rangle\langle\xi_0|$,
$\hat{t}_y(m)
\equiv-i|\xi_0\rangle\langle\xi_1|+i|\xi_1\rangle\langle\xi_0|$,
and 
$\hat{t}_z(m)
\equiv|\xi_0\rangle\langle\xi_0|-|\xi_1\rangle\langle\xi_1|$.
Any local operator on site $m$ is written
as $\hat{a}(m)=\sum_{\alpha=x,y,z}c_{\alpha,m}\hat{t}_\alpha(m)$.
By some calculation, we can show
$C(\hat{t}_\alpha(l),\hat{t}_\beta(l'),|\psi_1\rangle)
\le\sqrt{4\lambda_0\lambda_1}$
for $\alpha,\beta=x,y,z$ and $l'\in Z_1-l$.
Since 
$E(l)\equiv-\lambda_0\log_2\lambda_0-\lambda_1\log_2\lambda_1
\ge2\min(\lambda_0,\lambda_1)=2\lambda_1$,
we obtain
$C(\hat{t}_\alpha(l),\hat{t}_\beta(l'),|\psi_1\rangle)
\le\sqrt{4\lambda_0\lambda_1}
\le\sqrt{4\lambda_1}
\le\sqrt{2E(l)}$.
Let us assume that 
$\sum_{l\in Z_1}\sqrt{2E(l)}=o(N)$.
Then,
\begin{eqnarray*}
C(\hat{A},\hat{A},|\psi_1\rangle)
&=&
\sum_{l\in Z_1}
\sum_{l'\in Z_1-l}
\sum_{\alpha,\beta}
c_{\alpha,l}c_{\beta,l'}
C(\hat{t}_\alpha(l),\hat{t}_\beta(l'),|\psi_1\rangle)\\
&&+\sum_{l\in Z_1}
\sum_{\alpha,\beta}
c_{\alpha,l}c_{\beta,l}
C(\hat{t}_\alpha(l),\hat{t}_\beta(l),|\psi_1\rangle)\\
&\le&\big(|Z_1|-1\big)\sum_{l\in Z_1}\sqrt{2E(l)}+|Z_1|\\
&=&o(N^2),
\end{eqnarray*}
which means that $|\psi_1\rangle$ has $p<2$.
Since it contradicts to the assumption,
we obtain
$\sum_{l\in Z_1}\sqrt{2E(l)}=O(N)$.
Since $0\le E(l)\le 1$, this means that
the number of $l\in Z_1$ such that
$E(l)=O(1)$ is $O(N)$.



\begin{thebibliography}{00}
\bibitem{robustness}
Since the logarithmic
robustness is larger than the relative entropy entanglement
for pure states~\cite{Hayashi}, 
the logarithmic robustness is also 
large if $p=2$.
\end{thebibliography}
\end{document}